\RequirePackage{fix-cm}
\documentclass[smallcondensed]{svjour3ujm}     
\smartqed  
\usepackage{graphicx}
\usepackage[authoryear]{natbib}
\newcommand\bq{\par\begin{quote}}
\newcommand\eq{\par\end{quote}}
\newcommand{\ket}[1]{|#1\rangle}
\journalname{Foundations of Science}
\begin{document}

\title{A QBist Ontology
}


\author{U.J. Mohrhoff}

\institute{Ulrich Mohrhoff \at
              Sri Aurobindo International Centre of Education\\
              605002 Pondicherry India\\
              \email{ujm@auromail.net}}
\date{Received: date / Accepted: date}
\maketitle
\begin{abstract}
This paper puts forward an ontology that is indebted to QBism, Kant, Bohr, Schr\"odinger, the philosophy of the Upanishads, and the evolutionary philosophy of Sri Aurobindo. Central to it is that reality is relative to consciousness or experience. Instead of a single mind-independent reality, there are different poises of consciousness, including a consciousness to which ``we are all really only various aspects of the One'' (Schr\"odinger). This ontology helps clear up unresolved issues in the philosophy of science, such as arise from the reification of either instruments or calculational tools, or from a disregard of the universal context of science, which is human experience. It further helps clear up unresolved issues in the philosophy of mind, among them the problem of intentionality and the dilemma posed by the mutual inclusion of self and world (Husserl's paradox of human subjectivity).\keywords{Bohr \and Consciousness \and Experience \and Kant \and Mind-body problem \and QBism \and Schr\"odinger \and Upanishads}
\end{abstract}

\section{Introduction}
I find myself in excellent company. Like the distinguished philosopher Hilary Putnam (and others certainly), I keep changing my mind.%
\footnote{\emph{The Philosophical Lexicon} \citep{hilary} contains the following entry:\par
\leftskip=15pt\textbf{hilary}, n. (\emph{from} hilary term) A very brief but significant period in the intellectual career of a distinguished philosopher. ``Oh, that's what I thought three or four hilaries ago.''}
The definite statement of the previous version of what started out in 2000 as the ``Pondicherry interpretation'' of quantum mechanics \citep{Mohrhoff2000} [in the form of a comment on the ``Ithaca interpretation'' by David \citet{Mermin98}] was published in 2017 in this journal \citep{Mohrhoff2017}. There I attributed the necessity of distinguishing between a non-contextual classical domain and a contextual quantum domain to the difference between the \emph{manifested} world and its \emph{manifestation}. I envisioned the manifestation of the world as a progressive transition from the undifferentiated unity of a single ontological substance to a world which allows itself to be described in the object-oriented language of classical discourse. Subatomic particles, non-visualizable atoms, and partly visualizable molecules mark the stages of this transition. Instead of being constituent parts of the manifested world, they play an instrumental role in its manifestation.

Two questions arise in this framework. Firstly, how are we to describe the intermediate stages of the transition, at which the definiteness of properties and the distinguishability of objects are not as yet fully realized? My answer to this question was (and still is) that whatever is not \emph{intrinsically} definite can only be described in terms of probability distributions over what \emph{is} intrinsically definite, to wit: the possible outcomes of measurements. What is instrumental in the manifestation of the world can only be described in terms of correlations between events that happen (or could happen) in the manifested world. This goes a long way towards explaining why the general theoretical framework of contemporary physics is a probability calculus, and why the events to which it serves to assign probabilities are measurement outcomes.

And secondly, what accounts for the intrinsic definiteness of measurement outcomes? It is not enough to argue, as I did at the time, that the measurement apparatus is needed to \emph{realize} or \emph{define} the values that quantum observables can possess, or to make properties available for attribution to quantum objects. The measurement apparatus is also needed to \emph{indicate} the actually possessed value of a quantum observable or the actually possessed property of a quantum object, and no property or value can be indicated in the absence of a sentient observer to whom it is indicated. The reason why the properties of the manifested world are intrinsically definite is that the manifested world is an \emph{experienced} world. It is manifested \emph{to us}.

What made me come round to seeing that there is \emph{no difference} between observations qua measurement outcomes and observations qua (subjective) experiences, which are intrinsically definite, was a new and radically epistemic interpretation of quantum mechanics, launched early in the 21st Century by \citet{CFS2002}. Initially conceived as a generalized personalist Bayesian theory of probability called ``Quantum Bayesianism,'' it has since been re-branded as ``QBism,'' the term \citet{MerminQBnotCop} prefers, considering it ``as big a break with 20th century ways of thinking about science as Cubism was with 19th century ways of thinking about art.'' The big break lies not in the emphasis that the mathematical apparatus of quantum mechanics is a probability calculus but in this \emph{plus} a radically subjective Bayesian interpretation of probability \emph{plus} a radically subjective interpretation of the events to which, and on the basis of which, probabilities are assigned by ``users'' (of quantum mechanics) or ``agents'' (in a quantum world).%
\footnote{While Fuchs and Schack prefer the term ``agent,'' Mermin prefers ``user,'' to emphasize that QBists regard quantum mechanics as a ``user's manual'' \citep{MerminQBnotCop}.}

To drive home the ultimate context of empirical science, which is human experience, QBists emphasize the role of the \emph{individual} subject in constructing ``a common body of reality'' \citep{FMS2014}. At first experience is not \emph{ours}; it is \emph{yours} and \emph{mine}. It \emph{becomes} ours, and the process by which it becomes ours is communication. Science is seen as ``a collaborative human effort to find, through our individual actions on the world and our verbal communications with each other, a model for what is common to all of our privately constructed external worlds'' \citep{MerminQBnotCop}. 

I am indebted to QBism for another reason: it made me look more closely at the original writings of Niels Bohr and discover (i)~that the different versions of
the so-called Copenhagen interpretation misrepresent Bohr's views in various
degrees, and (ii)~that Bohr himself came tantalizingly close to adopting a QBist stance \citep{UJMBohrObIrrev}. The principal differences between his views and QBism are attributable to the fact that Bohr wrote before interpreting quantum mechanics became a growth industry, while QBism emerged in reaction to an ever-growing number of attempts at averting the ``disaster of objectification'' (van Fraassen, \citeyear{vF1990}) caused by the reification of (almost always) unitarily evolving quantum states.%
\footnote{Earlier invocations of consciousness, starting with von \citet{vN} and \citet{LonBau}, were unsuccessful attempts to avert this disaster. These should not be mixed up with QBism.}

As my engagement with QBism made me realize that the incontestable definiteness of human sensory experience alone can justify the definiteness of observations, so my subsequent engagement with the Bohr canon convinced me that this is what Bohr himself had been trying to say \citep{UJMBohrObIrrev}. When Bohr stressed the ``perfectly objective character'' of the quantum-mechanical description of atomic phenomena, it was ``in the sense that no explicit reference is made to any \emph{individual} observer'' \citep[p.~128, emphasis added]{BCW10}. It was never in the sense that no reference was made to the \emph{community} of observers. Reference to individual observers was implicit in all such ``remarkably QBist-sounding pronouncements'' \citep{MerminQBnotCop} as the following: ``in our description of nature the purpose is not to disclose the real essence of the phenomena but only to track down, so far as it is possible, relations between the manifold aspects of our experience'' \citep[p.~296]{BCW6}. It is \emph{our} experience because it is communicable, and it is communicable because it is expressible in terms of classical concepts. To Bohr, classical concepts are classical not because they are proprietary to classical physics but because we all know what they mean, inasmuch as their meanings are rooted in what we all have in common, to wit: the spatiotemporal structure of human sensory experience and the logical or grammatical structure of human thought or language \citep{UJMBohrObIrrev}.

With regard to the roots of objectivity, then, QBism is on the same page with Bohr. In particular, they both emphasize the importance of ordinary language in constructing an objective reality.%
\footnote{Bohr insisted that ``all well-defined experimental evidence, even if it cannot be analysed in terms of classical physics, must be expressed in ordinary language'' (\citeyear[p.~355]{BCW7}), i.e., ``plain language suitably refined by the usual physical terminology'' (\citeyear[p.~390]{BCW7}) or ``conveniently supplemented with terminology of classical physics'' (\citeyear[p.~277]{BCW10}).}
But this also means that neither evinces a desire to go beyond the intersubjective agreement that is achieved through communication. The question of whether we only experience our respective private worlds or actually experience one and the same world is left open.

\citet[p.~69]{SchrWhatIsReal}, too, who according to \citet{FMS2014} took ``a QBist view'' of science, would have agreed with QBism about the role that language plays in establishing the correspondence between, as he put it, ``the content of any one sphere of consciousness and any other, so far as the external world is concerned.'' What does establish it for him ``is \emph{language}, including everything in the way of expression, gesture, taking hold of another person, pointing with one's finger and so forth, though none of this breaks through that inexorable, absolute division between spheres of consciousness'' (original emphasis). But Schr\"odinger also made it clear that establishing something is not the same as accounting for it. To him, the agreement between the content of my sphere of consciousness and that of yours was not rationally comprehensible: ``In order to grasp it we are reduced to two irrational, mystical hypotheses,'' he wrote (\citeyear[p.~106]{SchrWhatIsReal}). 

One of these was ``the so-called hypothesis of the real external world.'' According to it there is ``a real world of bodies which are the causes of sense-impressions and produce roughly the same impression on everybody'' (\citeyear[p.~67--68]{SchrWhatIsReal}). Schr\"odinger left no room for uncertainty about what he thought of it: to invoke the existence of such a world ``is not to give an explanation at all; it is simply to state the matter in different words. In fact, it means laying a completely useless burden on the understanding''---the burden of making sense of the relation between our experiences and an empirically inaccessible ``real'' world. While we can use language to compare our respective experiences, we have no way of comparing our experiences with this world of bodies presumed to contain the causes of our experiences. The other hypothesis, which he endorsed, was that ``we are all really only various aspects of the One'' \citep[p.~106]{SchrWhatIsReal}. It will be the subject of Sec.~\ref{sec:Schr-take}.
 
Since previously I have been critical of QBism (e.g., Mohrhoff, \citeyear{UJM-QBismAppraisal}), the title of the present paper may seem misleading or inappropriate. One referee wondered if it was meant ironically. It was not. Apart from QBism's grounding of the definiteness of measurement outcomes in human experience, there is much to commend it. I agree with QBism that the mathematical formalism of quantum mechanics is (and only makes sense as) a calculus for assigning probabilities to measurement outcomes on the basis of information derived from measurement outcomes.%
\footnote{In \emph{The Ashgate Companion to Contemporary Philosophy of Physics} \citep{Wallace2008}, a distinction is made between  the ``bare quantum formalism,'' which is regarded as ``an elegant piece of mathematics \dots\ prior to any notion of probability, measurement etc.,'' and the ``quantum algorithm,'' which is looked upon as ``an ill-defined and unattractive mess.'' In truth, there is no such thing as a \emph{bare} quantum formalism. Every single axiom of any axiomatization whatsoever of the quantum theory only makes sense as a feature of a probability calculus \citep[Sec.~20.2]{Mohrhoff-book}.}
 ``If quantum theory is so closely allied with probability theory,'' \citet{Fuchs2010} once asked, ``why is it not written in a language that \emph{starts} with probability, rather than a language that ends with it?'' QBism has come close to writing it in just such a language. All that quantum mechanics needs is the Born rule, provided that it is formulated in terms of positive-operator-valued probability measures that are informationally complete, rather than in terms of the standard projector-valued measures. If this is done, one finds that ``[t]he Born Rule is nothing but a kind of Quantum Law of Total Probability!  No complex amplitudes, no operators---only probabilities in, and probabilities out'' \citep{Fuchs2010}.%
\footnote{The most general form of the Born rule is too unwieldy to replace the amplitudes and operators of the standard formalism. A simple and elegant form is obtained whenever SIC (\emph{symmetric} informationally complete) measurements can be used, but proofs of their existence are elusive. As of May 2017, such proofs have been found for all dimensions up to 151, and for a few others up to 323 \citep{Fuchs_Notwithstanding}. The mood of the QBist community nevertheless is that a SIC measurement should exist for every finite dimension.}

QBism is not just another interpretation of quantum mechanics. It is a view of science that ``puts the scientist back into science'' \citep{MerminNatureQBism}. The insight that the universal context of science is human experience is as old as Kant's theory of science. More recently \citet{Schilling}, in a paper presented during an AAAS meeting in 1955, contrasted popular belief about science as ``a sort of intellectual machine, which \dots\ inevitably grinds out ultimate truth in sequential steps'' with ``science as lived by its practitioners,'' and concluded that 
\bq
[w]e have come to realize, as perhaps no scientists before us ever have, that the human observer or explorer and his experience are integral and determinative
parts of whatever world he is studying.
\eq
But it is the QBist solution to one of the most pressing conundrums in the philosophy of science---the (small) measurement problem%
\footnote{\label{note_bigmp}A distinction has been drawn between the ``big'' measurement problem and the ``small'' one \citep{Pitowsky2006}. The big one, which calls for a dynamic account of how measurements come to have outcomes, only arises in ontologies that reify quantum states. The small measurement problem calls for an explanation of why we do not experience superpositions of macroscopically distinct states.}%
---that has brought home to me that the world studied by scientists cannot be detached from the scientists by whom it is studied. Only the character of human sensory experience can account for the definiteness of measurement outcomes or the fact that we do not experience superpositions of macroscopically distinct states. 

There are other currents of thought that have flowed into the ontology presented in these pages, least but not last my understanding that the difference between classical and quantum is rooted in the difference between the manifested world and its manifestation. There are Kant and Bohr, there is Schr\"odinger's solution (inspired by the philosophy of the Upanishads) to what he called the ``arithmetical paradox'' \citep{SchrLifeMindMatterAP}, and there is \citet{SA-Kena,SA-Isha}, who is arguably the most qualified modern interpreter of Upanishadic thought. But it is QBism that brought these currents together in the realization that the manifestation of the world does not take place in a mindless vacuum. The manifested world is an \emph{experienced} world. It is manifested \emph{to us}. In Upanishadic terms, the One (as a single conscious substance) manifests the world to itself, and therefore to us who are but ``various aspects of the One.'' 

QBism, as most QBists will agree, is a work in progress.%
\footnote{When I wrote to Mermin that QBism appears to mean different things to different QBists---although this may only be an appearance---he replied (on 5/29/2019): ``No, it's not just an appearance.  QBism is a work in progress, still under construction. Welcome to the crew.''}
Certain claims that have been made by QBists appear to be inconsistent with the general drift of QBism, and some of them are discussed in an appendix. Particularly unfortunate are the inconsistencies that arise from an ambiguous use of the expression ``external world.'' One possible meaning for this expression is a collectively constructed ``common body of reality'' \citep{FMS2014} or ``a model for what is common to all of our privately constructed external worlds'' \citep{MerminQBnotCop}. Another possible meaning is invoked in the following passage, in which \citet{FS2004} refer to ``the world external to'' the agent as ``the world as it is without agents'': 
\bq
The agent, through the process of quantum measurement stimulates the world external to himself. The world, in return, stimulates a response in the agent that is quantified by a change in his beliefs---i.e., by a change from a prior to a posterior quantum state. Somewhere in the structure of those belief changes lies quantum theory's most direct statement about what we believe of the world as it is without agents.
\eq
While the first meaning depends on a community of individual agents, the second is independent of agents. The tension between the two senses is particularly jarring in such statements as the following, which occur not infrequently  in the QBist literature. \citet{Mermin2019} writes: ``The world acts on me, inducing the private experiences out of which I build my understanding of my own world.'' If we build our understandings of our respective private worlds---and therefore our common body of reality---out of the private experiences the world has induced in us, the world that induces experiences in us cannot be uncritically identified with the shared external reality we have constructed out of our private experiences.

How the world that we construct on the basis of our experiences is related to whatever it is that induces experiences in us, has been a central issue throughout the history of philosophy. It can be stated in the form of a dilemma, whose two horns are (i) the existence \emph{of} consciousness \emph{in} what appears to be a material world and (ii) the existence \emph{in} consciousness \emph{of} what appears to be a material world. \citet[p.~178]{Husserl} referred to it as ``the paradox of human subjectivity: being a subject for the world and at the same time being an object in the world.'' 

Taking hold of the dilemma's objective horn (and focusing on perception), we gain an understanding of certain causal chains leading from external objects to firing patterns in brains, but then we find ourselves stymied by the notorious ``explanatory gap'' \citep{Levine} between neural processes and conscious experience. Here is how the issue has been presented by \citet[pp. 7--8]{Putnam_MFR}: 
\bq
How does the familiar explanation of what happens when I ``see something red'' go? The light strikes the object (say, a sweater), and is reflected to my eye. There is an image on the retina\dots. There are resultant nerve impulses\dots. There are events in the brain, some of which we understand thanks to the work of Hubel and Wiesel, David Marr, and others. And then---this is the mysterious part---there is somehow a ``sense datum'' or a ``raw feel.'' This is an explanation? An ``explanation'' that involves connections of a kind we do not understand at all \dots\ and concerning which we have not even the sketch of a theory is an explanation through something more obscure than the phenomenon to be explained.
\eq
Taking hold of the dilemma's subjective horn, we gain an understanding of the general structure of empirical knowledge, its dependence on the logical structure of rational thought and the spatiotemporal structure of (human) perceptual experience, but then we find ourselves stymied by the same explanatory gap approached from the other side: the brains studied by neuroscience are themselves objects of perceptual experience, and objects of perceptual experience cannot be uncritically identified with whatever it is that causes perceptual experience.

Most philosophers have found the paradox of subjectivity intolerable; hence the common attempt to eliminate one or the other of our two selves---the transcendental self \emph{for} which the world exists, or the empirical self which exists in the world, as an aspect or attribute of a physical body. The common reaction in our own day is to eliminate subjectivity altogether by some kind of physicalist reduction. For the purpose of developing a QBist ontology, only the approach which starts with the transcendental self is available.%
\footnote{Nor would direct realism as defended by \citet{Searle2004} fit the bill \citep{Mohrhoff-B4B}.}
It will be taken up in Sec.~\ref{sec:Kant} with an outline of Kant's theory of science.%
\footnote{I second von Weizs\"acker's (\citeyear{vW}) recommendation: ``Those who really want to understand contemporary physics---i.e., not only to apply physics in practice but also to make it transparent---will find it useful, even indispensable at a certain stage, to think through Kant's theory of science.''}

Quantum mechanics, too, has given rise to two mutually incompatible approaches: one fundamentally philosophical, the other essentially mathematical; one spearheaded by Niels Bohr, the other set in motion by John von Neumann. Here, too, only one---the first---is compatible with QBism. Bohr's philosophy of quantum mechanics, therefore, is the subject of Sec.~\ref{sec:Bohr}. ``From our present standpoint,'' \citet[p. ~157]{BCW10} wrote in a strikingly Kantian vein, ``physics is to be regarded not so much as the study of something \emph{a priori} given, but rather as the development of methods for ordering and surveying human experience.'' What Bohr added to Kant's theory of science was his insight that empirical knowledge is not necessarily limited to what is \emph{directly} accessible to our senses, and that, therefore, it does not have to be \emph{solely} a knowledge of sense impressions organized into objects. It can also be a knowledge of properties that (i)~are defined by experimental arrangements (which \emph{are} directly accessible to our senses), and that (ii)~only exist if their presence is indicated by the results of actual experiments.%
\footnote{Affinities between Bohr and Kant have been noted by a number of scholars \citep{Bitbol2010,Bitboletal,Brock,Chevalley94,Cuffaro,Falkenburg2009,FF94,Folse94,Honner1982,Hooker94,Kaiser1992,MacKinnon2012}.}

While we owe to Bohr the insight that the \emph{properties} of quantum systems are contextual, he was less explicit about the contextuality of the carriers of these properties. The latter was brought into focus first by Schr\"odinger and more systematically half a century later by Ulfbeck and (Aage) Bohr and by Brigitte Falkenburg, as will be discussed in Sec.~\ref{sec:pm}. According to \citet[pp. 205--206, original emphasis]{Falkenburg},
\bq
only the experimental context (and our ways of conceiving of it in classical terms) makes it possible to talk in a sloppy way of \emph{quantum objects}.\dots. Bare quantum ``objects'' \dots\ are only individuated by the experimental apparatus in which they are measured or the concrete quantum phenomenon to which they belong.
\eq
The overall conclusion of Secs.~\ref{sec:Bohr} and~\ref{sec:pm} is that if Kant did not rule out the elision of the experiencing and thinking subject, quantum mechanics certainly did.

If atoms and subatomic particles owe their properties to the experimental conditions under which they are observed, how can the experimental apparatus (or any other macroscopic object) be related to the atoms and subatomic particles of which it is commonly said to be composed? What is immediately given us in visual experience is not things but shapes, and quantum mechanics makes it possible to understand how the shapes we attribute to things are manifested (to us, in our experience). In this manifestation, atoms and subatomic particles play instrumental roles. This will be shown in Sec.~\ref{sec:shapes}. Moreover, because the shapes of things can be accounted for in terms of spatial relations between (ultimately) formless relata, we can invoke the principle of the identity of indiscernibles and regard the relations that constitute all forms not as relations between a multitude of relata but as the self-relations of a single relatum. Subsequently, in Sec.~\ref{sec:theOne},
 it will be argued that ``the miraculous identity of particles of the same type'' \citep{Misneretal} is actually a feature of \emph{all} particles, not only of particles of the same type. 

So far the question of whether we only experience our respective private worlds or actually experience one and the same world has remained open. But if the spatial relations that make up the formal aspect of quantum objects are reflexive and thus entertained by a single relatum, then there has to be more to the agreement between ``the content of any one sphere of consciousness and any other'' \citep[p.~69]{SchrWhatIsReal} than is warranted by language. This is where Schr\"odinger's appeal to the Upanishads comes in. What explains the agreement is that ``we are all really only various aspects of the One'' (p.~106)---the one Self of the Upanishads, the Ultimate Subject from which we are separated by a veil of self-oblivion. The same veil, according to the Upanishads, also prevents us from perceiving the Ultimate Object, as well as its fundamental identity with the Ultimate Subject. It prevents us from perceiving that the world is something that the One (qua Ultimate Object) manifests to itself (qua Ultimate Subject)---and therefore to us who are but ``various aspects of the One.'' But if at bottom we are all the same subject, we have to conceive of two poises of consciousness or modes of awareness, one in which the One manifests the world to itself \emph{aperspectivally}, as if experienced from no particular location or from everywhere at once, and one in which the One manifests the world to itself \emph{perspectivally}, as if experienced by a multitude of subjects from a multitude of locations. This takes the edge of Husserl's paradox of human subjectivity. As situated aspects of the One, we appear to be objects in the world; as the One, we both constitute and contain the world. This is the gist of Sec.~\ref{sec:Schr-take}. 

An idea that has been dominant throughout history is that neural representations are encodings. Section~\ref{sec:synergy} addresses the nature of the interpreter of the encoded information, the basis of its interpretive capacity, and the problem of intentionality, which may be the most pressing issue in the contemporary philosophy of mind. 

The main plot of the self-manifestation of the One is a cycle of self-concealment (involution) and self-discovery (evolution). Section~\ref{sec:evolution} outlines the steps by which the stage for the adventure of evolution was set. It also addresses the reason why the One would want to enter such a cycle, considering the pain and suffering that (in hindsight) it entails. Section \ref{sec:future}, finally addresses the part that physics plays in all this, touches on the conundrum of free will, and speculates about the future of evolution.

\section{Beyond correspondence}\label{sec:Kant}
The oldest  attempts to bridge the explanatory gap invoke some kind of correspondence. A correspondence theory (of truth) is a theory that frames the relation between the world we experience (or construct on the basis of our experiences) and the reality that induces experiences in us, as some kind of correspondence. The oldest known such theory, which endured for approximately 2,000 years, is the one that ancient and medieval philosophers have attributed to Aristotle. It holds that the relation between the \emph{phantasm} and the external object, by virtue of which the \emph{phantasm} represents the external object to the mind, is literally a relation of similarity. 

To 17th-Century thinkers like Descartes and Locke, it still seemed to pose no difficulty to conceive of perceived sizes and shapes as similar to real sizes and shapes. That a perceived color should be similar to color in the external world was a more questionable proposition (as it had been for Aristotle). Locke and Descartes therefore distinguished between ``primary qualities,'' which were independent of the perceiving subject, and ``secondary qualities,'' which bore no similarities to sensations but had the power to produce sensations in the perceiving subject. Eventually, though, thinking of perceived sizes and shapes as similar to real sizes and shapes proved to be no less questionable than the proposition that color sensations are similar to colors in the external world. (Berkeley made it clear that to ask whether a table is the same size and shape as my mental image of it was to ask an absurd question.)

Enter Kant, whose \emph{Critique of Pure Reason} ranks as one of the most influential philosophical works of all time. Its appearance marks the end of the modern period and the beginning of something entirely new \citep{KantSEP}. In it Kant starts out by saying that \emph{all} qualities are secondary in Locke's sense. Nothing of what we say about an object describes the object as it is in itself, independently of how it affects us. Nor does Kant stop at saying that if I see a desk, there is a thing-in-itself that has the power to appear as a desk, and if I see a chair in front of the desk, there is another thing-in-itself that has the power to appear as a chair. For Kant, there is only \emph{one} thing-in-itself, an empirically inaccessible reality that has the power to affect us in such a way that we have the sensations that we do, and that we are able to ``work up the raw material of sensible impressions into a cognition of objects'' \cite[p.~136]{KantCPR}. 

Kant owes his fame in large part to his successful navigation between the Scylla of commonsense realism and the Charybdis of idealism. What allowed him to steer clear of both horns of \emph{this} dilemma was a dramatic change of strategy. Instead of trying to formulate a metaphysical picture of the world consistent with Newtonian mechanics, as he had done during the pre-critical period of his philosophy, he inquired into the cognitive pre-conditions of the possibility of natural science. ``Look,'' he might have said, ``natural science exists. We have Newton's laws of motion, the law of gravity, and various other laws. So let us find out what must be the case \emph{on our part} so that we can have such a thing as natural science.'' He did not presume to inquire into the \emph{metaphysical} underpinnings of this fact. He resolved to determine its \emph{cognitive} underpinnings. The position taken by him was that the objective world is constructed by the human mind from sensory material that is passively received and concepts that owe their meanings to the logical structure of rational thought  and the spatiotemporal structure of human sensory experience. The most important of these concepts are \emph{substance} and \emph{causality}. 

The link between substance and logic was first forged by Aristotle. To Aristotle, a property was whatever could be the predicate of a logical subject, while a substance was something that could not be predicated of anything else. Substances, therefore, enjoyed independent existence, while properties owed their existence to being attributes of substances. \citet{Locke} subsequently distinguished between two conceptions of substance: (i)~a ``notion of pure substance in general'' and (ii)~``ideas of particular sorts of substance.'' The first was to him ``nothing but the supposed, but unknown, support of those qualities we find existing, which we imagine cannot subsist \emph{sine re substante}, without something to support them.'' In other words, it was something that the qualities we find existing do not need---``any more than the earth needs an elephant to rest upon,'' as \citet{RussellHistory} later put it. Substances in sense~(ii) are sometimes called ``Lockean substances.'' They are ``such combinations of simple ideas as are, by experience and observation of men's senses, taken notice of to exist together.'' They bring into play another function of the logical (or grammatical) subject: substance serves to combine different ``simple ideas'' in the same way (or the same sense) that a single subject serves to combine several predicates.

In Kant's theory, substance fulfills both functions: like a Lockean substance it serves to bundle sensible impressions in the manner in which a logical or grammatical subject bundles predicates, and like an Aristotelean substance it makes it possible for me to think of my sensible impressions as connected not in or by me but in or by an external object. This allows me to ignore the fact that the external object owes its existence largely to me, the thinking subject that ``work[s] up the raw material of sensible impressions into a cognition of objects'' \cite[p.~136]{KantCPR}.

The concept of causality likewise performs two functions. It allows me to think of impressions received at different times as connected in accordance with the logical or grammatical relation between an antecedent and a consequent, and it makes it possible for me to think of my successive perceptions as connected not in me, by my experiencing them, but objectively, as causes and effects in an external world. (Kant believed that another concept was needed to objectivize the temporal relation of simultaneity. In a relativistic world, in which simultaneity cannot be objectivized, no such concept is required.)

Furthermore, the possibility of thinking of my perceptions as a self-existent system of external objects requires that the connections be lawful. If sense impressions are to be perceptions of a particular kind of object (say, an elephant) they must be connected in an orderly fashion, according to a concept denoting a lawful concurrence of perceptions. And if sense impressions are to be perceptions of causally connected occurrences, like (say) lightning and thunder, they must fall under a causal law, according to which one perception necessitates the subsequent occurrence of another. To Kant, therefore, natural science was concerned with causal laws and Lockean substances.%
\footnote{Kant maintained that by establishing lawful causal relations, we also establish objective temporal relations. Although today we have a deeper understanding of the relationship between physical law and spacetime geometry, Kant was right in stressing the intimate relation between physical law and temporal relations. Because we cannot formulate a fundamental physical law without presupposing a particular geometry, there is no ``true'' geometry. As was stressed by \citet{Poincare}, ``[o]ne geometry cannot be more true than another, it can only be more convenient.'' For any physical theory, the best spacetime geometry to use is one that yields the simplest mathematical formulation of the theory. In Kant's time, this was Euclidean geometry plus Newtonian uniform time, today it is the Minkowski geometry of special relativity or the pseudo-Riemannian geometry of Einstein's theory of gravity.}

\section{Niels Bohr}\label{sec:Bohr}
The crucial premise of Kant's inquiry was that (i)~``space and time are only forms of sensible intuition,%
\footnote{``Intuition'' is the standard translation of the German word \emph{Anschauung}, which covers both visual perception and visual imagination. Perception is sensible, i.e., filled with actual sensations, imagination is not.}
and therefore only conditions of the existence of the things as appearances,'' that (ii)~``we have no concepts of the understanding and hence no elements for the cognition of things except insofar as an intuition can be given corresponding to these concepts,'' and that therefore (iii)~``we can have cognition of no object as a thing in itself, but only insofar as it is an object of sensible intuition, i.e. as an appearance'' \citep[p.~115]{KantCPR}. Niels Bohr could not have agreed more, insisting as Kant did that meaningful physical concepts have not only mathematical (quantifiable) but also visual (spatiotemporal) content.%
\footnote{\label{note_cc}Position and orientation are obviously of this kind. So are linear and angular momentum (which derive their meanings from the symmetry properties of space or the invariant behavior of closed systems under translations and rotations) as well as energy (which derives its meaning from the uniformity of time or the invariant behavior of closed systems under time translations).}
But  \citet[p.~217]{BCW6} also realized that ``the facts which are revealed to us by the quantum theory \dots\ lie outside the domain of our ordinary forms of perception.'' By ``our ordinary forms of perception'' Bohr meant
\bq
the conceptual structure upon which our customary ordering of our sense-impressions depends and our customary use of language is based. The basis of this ordering is, certainly, the possibility for recognition and comparison and accordingly the usual description of nature is characterized by the attempt to express all experience by stating the locations of material bodies and changes of location with time relative to a coordinate system defined in the traditional manner by means of measuring rods and clocks. \cite[pp. xxxv--xxxvi]{BCW10}
\eq
In other words, he meant the system of concepts that allows us to unambiguously identify (recognize and compare) objects in space and time. But this is the very system of concepts whose applicability Kant has shown to be a precondition of the possibility of objective knowledge, by which Kant meant knowledge of a (for all practical purposes subject-independent) system of objects.

What Kant did not anticipate was the possibility of an empirical knowledge that, while being obtained \emph{by means of} sense impressions organized into objects, was not a knowledge \emph{of} sense impressions organized into objects. Bohr realized that quantum mechanics was that kind of knowledge. What Bohr added to Kant's theory of science was his insight that empirical knowledge was not necessarily limited to what is \emph{directly} accessible to our senses, and that, therefore, it did not have to be \emph{solely} a knowledge of sense impressions organized into objects. It can also be a knowledge of properties that (i)~are \emph{defined} by experimental arrangements (which \emph{are} directly accessible to our senses), and that (ii)~actually exist only if their presence is indicated by the results of actual experiments. The click of a counter does not simply indicate the presence of something inside the region monitored by the counter. The counter \emph{defines} a region, and the click \emph{constitutes} the presence of something within it. Without the click, \emph{nothing} is there, and without the counter, there is no \emph{there}.

As long as the \emph{only} relevant context of empirical science was human experience (as it was for Kant), or as long as the reach of human sensory experience was potentially unlimited (as it was for Kant, Newton, and classical physics in general), the elision of the subject could be achieved: one could think and behave as if the objective world existed independently of perceiving and conceiving subjects. Having asserted that ``we can have cognition of no object as a thing in itself, but only insofar as it is an object of sensible intuition, i.e. as an appearance,'' \citet[p.~115]{KantCPR} could go on to affirm that
\bq
even if we cannot \emph{cognize} these same objects as things in themselves, we at least must be able to \emph{think} them as things in themselves. For otherwise there would follow the absurd proposition that there is an appearance without anything that appears. 
\eq
But now there was more than one relevant context. In classical physics, a single picture could accommodate all of the properties a system can have at any moment of time. When quantum mechanics came along, that all-encompassing picture fell apart. Unless certain experimental conditions obtained, it was impossible to picture the electron as following a trajectory (which was nevertheless a routine presupposition in setting up Stern--Gerlach experiments and in interpreting cloud-chamber photographs), and there was no way in which to apply the concept of position. And unless certain other, incompatible, experimental conditions obtained, it was impossible to picture the electron as a traveling wave (which was nevertheless a routine presupposition in interpreting the scattering of electrons by crystals), and there was no way in which to apply the concept of momentum. 

The implication was then (and still is) that the properties of atoms and subatomic particles owe their existence to the experimental conditions under which they are observed. The positions indicated by droplets forming track in a cloud chamber were the positions of a particle \emph{because} they were indicated by the droplets. The momenta indicated by the (imaginary) tangents on a track were the momenta of a particle  \emph{because} they were indicated by the track. As \citet{HeisenbergBahn} phrased it, \emph{``Die `Bahn' entsteht erst dadurch, daß wir sie beobachten''}---a particle's path only comes into being because we observe it. 

But if atoms and subatomic particles owe their properties to the experimental conditions under which they are observed, the experimental apparatus cannot owe its properties to the quantum-mechanical systems of which it is commonly said to be composed. And therefore neither the existence of microscopic objects (like atoms and subatomic particles) nor that of macroscopic objects (like the experimental apparatus) can be attributed to independently existing substances. The contextuality of the properties of microscopic objects thus implied that the elision of the subject could no longer be achieved.

This explains Bohr's concern about the objectivity of the quantum-mechanical description of atomic phenomena. To Kant, the ability to attribute properties to substances, and to connect them according to the principle of causality, were preconditions of the possibility of objective knowledge, and so they were to Bohr, who stressed that ``the objective character of the description in atomic physics depends on the detailed specification of the experimental conditions under which evidence is gained'' \cite[p.~215]{BCW10}. A detailed specification of these conditions can only be given if it is possible to describe them in terms of property-carrying substance conforming to causal laws. 

If ``the description of the experimental arrangement and the recording of observations must be given in plain language, suitably refined by usual physical terminology,'' it was because ``by the word experiment we can only mean a procedure about which we are able to communicate to others what we have done and what we have learned'' \citep[p.~128]{BCW10}. If we could not refer to property-carrying substances that conform to causal laws, we could not communicate to others what we have done and what we have learned. Hence when Bohr affirmed the ``perfectly objective character'' of the quantum-mechanical description of atomic phenomena, it was ``in the sense that no explicit reference is made to any \emph{individual} observer'' \cite[p.~128, emphasis added]{BCW10}. It was never in the sense that no reference was made to the community of observers. This is also obvious from the importance Bohr attached to communication in common language, which went so far as to allow him to define objectivity in terms of the latter: ``By objectivity we understand a description by means of a language common to all'' \citep[p. xxxvii]{BCW10}.

If ``the physical content of quantum mechanics is exhausted by its power to formulate statistical laws governing observations obtained under conditions specified in plain language,'' as it was to \citet[p. 159]{BCW10}, or if the quantum-mechanical formalism ``represents a purely symbolic scheme permitting only predictions \dots\ as to results obtainable under conditions specified by means of classical concepts,'' as it did to \citet[pp. 350--351]{BCW7},%
\footnote{To Bohr, as mentioned in the introduction, classical concepts are classical not because they are proprietary to classical physics but because we all know what they mean, inasmuch as their meanings are rooted in what we all have in common, to wit: the spatiotemporal structure of human sensory experience and the logical or grammatical structure of human thought or language \citep{UJMBohrObIrrev}. This includes the concepts mentioned in Note~\ref{note_cc} as well as those of substance and causality.}
the reason was twofold: (i)~Because ``the facts which are revealed to us by the quantum theory \dots\ lie outside the domain of our ordinary forms of perception'' \citep[p.~217]{BCW6}---in other words, because they are not accessible to direct sensory experience---they are not amenable to description in classical terms (i.e., by concepts that owe their meanings in part to the spatiotemporal structure of human sensory experience). The only possible description therefore is mathematical or statistical. (ii)~A statistical description must be given in terms of correlations between events that can be described in classical terms, i.e., events that are accessible to direct sensory experience. 

At first quantum theory seemed to require a radical departure from the classical universe of discourse staked out by Kant. After all, the correlations between a preparation and an observation---between ``the fixation of the external conditions, defining the initial state of the atomic system concerned and the character of the possible predictions as regards subsequent observable properties of that system'' and ``the test of such predictions'' \citep[p.~312]{BCW7}---could not be accounted for in classical terms. It is to Bohr's great merit that instead of discarding the classical universe of discourse he \emph{expanded} it, by adding an intrinsically unspeakable domain of quantum phenomena, which becomes speakable only in terms of statistical correlations between events that happen or can happen in the classical domain.

\section{Particle metaphysics}\label{sec:pm}
If quantum systems owe their attributes to being measured, may not the quantum systems themselves be constituted by the outcomes of measurements? At the beginning of the 21st Century---about the time QBism came on the scene---this suggestion was followed up by Ole Ulfbeck and Aage Bohr (\citeyear{UlfBohr}). Like Kant and later Niels Bohr, Ulfbeck and Bohr view space and time as ``a scene established for the ordering of experiences.'' Clicks given off by detectors belong to this scene. Particles traveling from a source to a detector and \emph{producing} clicks do not: ``the connection between source and counter is inherently non-local.'' While clicks are ``events in spacetime, belonging to the world of experience,'' there are no particles ``on the spacetime scene.'' The individual click is ``an event entirely beyond law.'' Genuinely fortuitous clicks, occurring by themselves, form ``the basic material that quantum mechanics deals with.'' These views echo misgivings about the particle concept that half a century earlier have been expressed by \citet[pp. 121-122, 131--132, original emphases]{SchrNGSH}:
\bq
When you observe a particle of a certain type, say an electron, now and here, this is to be regarded in principle as an \emph{isolated event}. Even if you do observe a similar particle a very short time later at a spot very near to the first, and even if you have every reason to assume a \emph{causal connection} between the first and the second observation, there is no true, unambiguous meaning in the assertion that it is \emph{the same} particle you have observed in the two cases\dots. It is beyond doubt that the question of ``sameness,'' of identity, really and truly has no meaning\dots. {We must not admit the possibility of continuous observation}. Observations are to be regarded as discrete, disconnected events. Between them there are gaps which we cannot fill in\dots. [I]t is better to regard a particle not as a permanent entity but as an instantaneous event. Sometimes these events form chains that give the illusion of permanent beings---but only in particular circumstances and only for an extremely short period of time in every single case.
\eq
We are accustomed to thinking of a measurement as ascertaining the possession, by a \emph{given} physical object or system, of one of several possible properties, which are determined by the setup. Specifically, we think of a position measurement as ascertaining the possession, by a \emph{given} particle, of one of several possible positions, which are defined by an array of detectors. In experimental particle physics, however, much the opposite is the case: while we have a large number of possible positions, each defined by a detector, no particle is given. What is given is clicks, and our task is to organize them into tracks indicating the successive positions of what we imagine to be persistent individuals---wrongly, as Schr\"odinger first and later Ulfbeck and Bohr have stressed. 

Although, for Ulfbeck and Bohr, there are neither electrons nor neutrons on the spacetime scene, ``clicks can be classified as electron clicks, neutron clicks, etc.'' It does, however, take more than one click to identify a click as an electron click or a neutron click. There has to be a sequence of clicks, and it must be possible to interpret each of the clicks as constituting the presence of the same kind of particle. Particle detectors are designed so that these conditions are satisfied FAPP, to use Bell's (\citeyear{Bell90}) famous acronym for ``for all practical purposes''.%
\footnote{The qualification ``FAPP'' becomes necessary whenever classical features are obtained by ignoring quantum features. In this case the quantum feature ignored is the uncertainty relation for position and momentum. Because the locations of the clicks are ``fuzzy'' enough, the correlations between the clicks make it possible FAPP to imagine a continuous path, to imagine a particle following it, and to imagine it causing the clicks that make up the observed track.}
A typical particle detector consists of zillions of individual detectors each monitoring a relatively small region of space or spacetime. Compared to the number of actual clicks elicited during each experimental run, the total number of possible clicks (or droplets, or whatever else makes up a track) is therefore enormous. This---along with the conservation of energy and momentum tempered by the uncertainty principles for energy and time and for momentum and position---makes it possible to observe tracks, to identify the type of particle associated with a given track, and to thereby determine the type to which the clicks constituting the track belong.%
\footnote{A track makes it possible to measure such quantities as its radius of curvature (in a magnetic field), the particle's time of flight, its kinetic energy, and/or its energy loss through ionization and excitation. Measuring three of these quantities is sufficient in principle to positively identify the particle type \citep{GS}. (The type of a neutral particle, which cannot be inferred directly from a track, can be inferred indirectly from the particle's interactions with charged particles, with the help of the conservation laws.)}
It must be emphasized, however, that there are no \emph{true} individuals; at best there are ``close enough'' individuals. While particle types and their possible interconversions are rigorously defined by theoretical/axiomatic particle physics, the particle types so defined are not rigorously individuated by the (only approximately continuous) tracks of experimental particle physics.

A diversity of particle concepts are in contemporary use. In a painstaking investigation of ``what exactly physicists mean when they talk about subatomic particles'' (to which this section owes its title), Brigitte \citet[p.~30]{Falkenburg} has shown that ``[t]he attribution of physical properties to subatomic particles is not based on one unified theory but on \emph{several incommensurable theories}'' (p.~324, original emphasis), for ``the operational, axiomatic, and referential aspects of physical concepts fall apart'' (p.~162). The axiomatic (chiefly field theoretical and group theoretical) aspect only refers to particle types. It refers neither to individual particles nor to the experiments by which particles are observed and investigated. The operational aspect is tied to the probabilistic or ensemble interpretation of quantum mechanics and to classical measurement laws, and
\bq
[t]he way in which quantum concepts refer can only be understood in \emph{contextual} terms. Quantum concepts refer to the properties of quantum phenomena which occur by means of a given experimental setup in a given physical context'' (p.~200, original emphasis).
\eq
By way of illustration, consider the scattering experiments which in modern particle physics are front and center. Each experiment begins with either a single particle beam aimed at a fixed target or two intersecting beams. Because the incoming particles are accelerated by making them pass through strong magnetic and electric fields over large distances, the laws of classical mechanics and classical electrodynamics are used to describe how particles of definite momentum $p$ are created. What happens next is described in quantum-mechanical terms. The particles are said to be prepared in a momentum state (the rigorous definition of which is a probability algorithm), and the mathematical picture of an incoming plane wave of wavelength $\lambda=2\pi\hbar/p$ is used. If the target is, say, a crystal, its structure (to be determined) is treated as belonging to the classical boundary conditions of the experiment being conducted, and the incoming plane wave is treated like a classical wave that gets diffracted by the target. After the scattering, either clicks or tracks are observed, depending on whether one is interested in spatial interference patterns or distributions of outgoing momenta and/or particle types. To determine the locations of the clicks or to measure the curvature and other properties of the tracks, one reverts to purely classical measurement methods. The underlying pragmatic ``philosophy'' has been aptly summed up by Wolfgang \citet{Ketterle} in a popular talk,  in which he said that after several years of practice one gets used to \emph{preparing waves and detecting particles}. 

While Bohr had stressed the contextuality of \emph{properties}, he was considerably less explicit about the contextuality of \emph{objects}. The latter was brought into focus by \citet[pp. 205--206, original emphasis]{Falkenburg}:
\bq 
only the experimental context (and our ways of conceiving of it in classical terms) makes it possible to talk in a sloppy way of \emph{quantum objects}\dots. Bare quantum ``objects'' \dots\ seem to be Lockean empirical substances, that is, collections of empirical properties which constantly go together. However, they are only individuated by the experimental apparatus in which they are measured or the concrete quantum phenomenon to which they belong. 
\eq
To Falkenburg (p. 339, original emphases), therefore, subatomic reality is a top-down construct:
\bq
The opposite \emph{bottom-up} explanation of the classical macroscopic world in terms of electrons, light quanta, quarks, and some other particles remains an empty promise. Any attempt at constructing a particle or field ontology gives rise to a \emph{non-relational} account%
\footnote{Physical reality, as conceived by \citet[p. 334]{Falkenburg}, is relational in three respects: (i)~it is context dependent, (ii)~it is defined relative to classical concepts, and (iii)~it is energy dependent such that (e.g.) the quark-antiquark and gluon content of nucleons increases with increasing scattering energies.}
of a subatomic reality made up of \emph{independent} substances and causal agents. But any known approach of this type is either at odds with the principles of relativistic quantum theory or with the assumption that quantum measurements give rise to actual events in a classical world. As long as the quantum measurement problem is unresolved, an independent quantum reality is simply not available.
\eq
What Falkenburg refers to at the end is the big measurement problem mentioned in Note~\ref{note_bigmp}, which is also known as the problem of objectification. Since insolubility theorems have been proved for this problem \citep{BLM96,Mittelstaedt98}, the actual and unconditional conclusion is that an independent quantum reality is not available---period. All of this goes to confirm that the objective world is the objectivized or objectivizable subset of our private and subjective experiences, the part we share and think about as external to ourselves.%
\footnote{I use the verb ``to objectivize'' and the noun ``objectivation'' for the process by which we construct our common external reality, and reserve the verb ``to objectify'' and the noun ``objectification'' for references to the inexplicable emergence of measurement outcomes from reified quantum states (as in ``the disaster of objectification'').}

Another indication that objective word is an objectivized world, projected outward and externalized by experiencing subjects, is that the objectivation has its limits. The contextuality of positions in particular---i.e., the dependence of attributable positions on the existence of position-defining detectors---limits the extent to which objective space can be conceived as ``intrinsically'' partitioned. This alone makes nonsense of the field-theoretic postulation of an independently existing spacetime manifold. Such a manifold is a useful, even indispensable tool for calculating scattering amplitudes, but as the basis for an ontological quantum reality, it is a recipe for disaster.

\section{The shapes of things}\label{sec:shapes}
If atoms and subatomic particles owe their properties to the experimental conditions under which they are observed, the following question arises: how is the experimental apparatus (or any other macroscopic object) related to the atoms and subatomic particles of which it is commonly said to be composed? 

What is primarily given to us in visual experience is not things but the shapes of things---not independently existing substances but shapes that we are generally able to attribute to re-identifiable Lockean substances. The parts that are primarily given to us in visual experience are therefore defined in spatial terms. In fact, until quantum mechanics came along, the parts of a material object were thought to be defined by boundaries acting in the manner of three-dimensional cookie cutters, while their forms were thought to be defined by boundaries separating their ``stuff-filled'' insides from their vacant outsides. Plato believed that the Universe could be described using five simple shapes---the solids named after him, four of which he believed to constitute the four elements Fire, Air, Water, and Earth---but he obviously could not account for the origin of these primordial shapes. The Greek atomists held that atoms came in an infinite variety of sizes and shapes, which likewise remained unaccounted for. \citet{Newton-nature} speculated ``that God in the Beginning form'd matter in solid, massy, hard, impenetrable, moveable Particles, of such Sizes and Figures, and with such other Properties, and in such Proportion to Space, as most conduced to the end for which he form'd them''---and he (Newton) left it at that.

The first to give a dynamic explanation of extension and impenetrability was the Serbo-Croatian polymath Roger Boscovich (Bo\v sko\-vi\'c). On the basis of his study of collisions, \citet{Bosco} arrived at an atomic theory in which matter was reduced to a dynamic system of relations between identical dimensionless ``points of force'' lacking mass and substantiality. His \textit{Theory of Natural Philosophy}, first published in 1758, was well known and remained influential for 150 years. His work inspired Faraday's lines of force, it advocated a relational view of space, and it accounted for the stability of objects in terms of equilibria between attractions and repulsions.%
\footnote{Here is how Henry Cavendish summarized the essentials of Boscovich's theory \cite[p.~51]{James04}: ``[M]atter does not consist of solid impenetrable particles as commonly supposed, but only of certain degrees of attraction and repulsion directed towards central points. They also suppose the action of two of these central points on each other alternately varies from repulsion to attraction numerous times as the distance increases. There is the utmost reason to think that both these phenomena are true, and they serve to account for many phenomena of nature which would otherwise be inexplicable.''}
So impressed was Friedrich Nietzsche by Boscovich's elimination of matter that he compared him to Copernicus.%
\footnote{``While Copernicus convinced us to believe, contrary to all our senses, that the earth does \emph{not} stand still, Boscovich taught us to renounce belief in the last bit of earth that \emph{did} `stand still,' the belief in `matter,' in the `material,' in the residual piece of earth and clump of an atom: it was the greatest triumph over the senses that the world had ever known'' \cite[p.~14]{NietzscheBGE}.}

While Boscovich arguably achieved the complete elimination of pre-existent forms, classical point mechanics, for which he laid the foundation, retained them in the shape of point masses. The first theory to provide a successful and complete dynamical reduction of form is quantum mechanics. If a quantum object has a form, this consists of indefinite spatial relations between component parts.%
\footnote{What is intended by the characterization of a physical quantity~$Q$, such as a relative position or orientation, as \emph{indefinite}, is that it can only be described by assigning  nontrivial probabilities to the (counterfactually) possible outcomes of an unperformed measurement of~$Q$.}
If it lacks components, it also lacks a form.%
\footnote{According to the current standard model of particles and forces, some quantum objects, including electrons and quarks, are fundamental in the sense of not being composed of other quantum objects. While such objects are often described as pointlike, this can only mean that they lack internal structure, which is another way of saying that they lack component parts.}
(Here I do not speak of \emph{individuated} quantum objects. I refer to such natural kinds as quarks, nucleons, nuclei, and atoms, which the axiomatics allows us to construct by increasingly approximative methods.)

The form of a bipartite quantum object---for instance, that of a hydrogen atom if the structure of its nucleus is ignored---consists of a single indefinite relative position. The time-independent forms of such an object can be classified in terms of the possible outcomes of the measurements of three quantities: the object's energy, its total angular momentum, and one component of its angular momentum. The form of a quantum object with $N$ components ``exists'' in a configuration space of $3\times N$ dimensions and consists of $N\times(N{-}1)/2$ indefinite relative positions. The abstract forms of nucleons, nuclei, atoms, and molecules ``exist'' in probability spaces of increasingly higher dimensions. At the molecular level of complexity, a different kind of form comes into (abstract) being: a 3-dimensional form that can be visualized, and this not merely as a distribution over a probability space, to wit: the spatial arrangement of the atoms constituting a molecule.%
\footnote{What contributes to making these configurations visualizable is that the indefiniteness of the distance $d$ between any pair of bonded atoms, as measured by the standard deviation of the corresponding probability distribution, is significantly smaller in general than the mean value of~$d$.}
If there is a quantum-classical boundary, it is molecules that straddle it. On the classical side are their atomic configurations, which change slowly, with electron wave functions following adiabatically.%
\footnote{Only molecules consisting of very few atoms are known to occur in energy and angular momentum eigenstates~\cite[p.~99]{Joosetal2003}.}

But if the shapes of quantum objects---including molecules, and eventually everything we tend to think of as ``made from'' atoms and molecules---can be accounted for in terms of spatial relations (relative positions and orientations) between (ultimately) \emph{formless} relata, then we can invoke the principle of the identity of indiscernibles and regard these spatial relations not as relations between a multitude of relata but as the \emph{self-relations of a single relatum}.

\section{The One}\label{sec:theOne}
Just as quantum mechanics is inconsistent with set-theoretic thinking about objective space---i.e., as ``a Many that allows itself to be thought of as a One,'' which was Cantor's definition of a set---so quantum mechanics is inconsistent with set-theoretic thinking about particles. Objective space is more appropriately conceived as a One that allows itself to be thought of as a Many \emph{up to a point}, and so is the multiplicity of particles \citep{Mohrhoff2017}. The multiplicity of individuated quantum objects is obviously limited by the experimental conditions on which the individuation of quantum objects depends, and the multiplicity of non-individuated quantum objects (e.g., the multiplicity of electrons in an atom) is readily conceivable as the result of a process by which a One enters into reflexive relations (i.e., relations with itself).

It is not a new idea that particles \emph{of the same type} are numerically identical, in the sense of being (multiple aspects of) one and the same thing. In his Nobel lecture, Feynman recalled: ``I received a telephone call one day at the graduate college at Princeton from Professor Wheeler, in which he said, `Feynman, I know why all electrons have the same charge and the same mass.' `Why?' `Because, they are all the same electron!'\,'' The present claim, on the other hand, is that ``the miraculous identity of particles of the same type,'' which according to \citet{Misneretal} is to be regarded ``not as a triviality, but as a central mystery of physics,'' need not be confined to particles of the same type. 

To see why \emph{all} fundamental particles, even those belonging to different types, can (and should) be regarded as numerically identical, consider, first, an elastic scattering event featuring a pair of incoming particles ``in'' the states $\ket 1$ and $\ket 2$ and a pair of outgoing particles ``in'' the states $\ket A$ and~$\ket B$.%
\footnote{For brevity's sake the customary phraseology of ``being \emph{in}'' such and such a state is used in lieu of the more correct mouthful.}
If the two particles are of the same type, we can say that initially there are two things with the respective properties 1 and~2, and that subsequently there are two things with the respective properties $A$ and~$B$, but we cannot say that there are two separate \emph{enduring} things. And if there are no enduring things, one readily agrees with Schr\"odinger \citep{BitbolQMCR} and \citet{LadyRoss} that there are no \emph{things}. What we can say is that there is \emph{one} thing, which is initially observed to have the properties 1 \emph{and}~2 (e.g., being here as well as being there, or moving Eastward as well as moving Westward), and which is subsequently observed to have the properties $A$ \emph{and}~$B$.

Next assume that the two particles are of different types $\alpha$ and $\beta$, that the initial states are $\ket{\alpha,1}$ and $\ket{\beta,2}$, and that the final states are $\ket{\alpha,A}$ and $\ket{\beta,B}$. What we can be sure of in this case is that initially there are two things each with a pair of properties ($\alpha,1$ and $\beta,2$), and that subsequently there are again two things each with a pair of properties ($\alpha,A$ and $\beta,B$). While it is customary to say in this case that we are dealing with two \emph{enduring} things, this cannot be advanced as the reason \emph{why} we observe a particle of type $\alpha$ and a particle of type $\beta$ both initially and in the end. On the contrary, it is the fact that we observe a particle of type $\alpha$ and a particle of type $\beta$ both initially and at the end that we are in a position to entertain the belief that we are dealing with two enduring things. This belief has the same FAPP status as the belief that the clicks that make up a track indicate the presence of the same enduring individual. What remains rigorously assertible is that there is \emph{one} thing, which is initially observed in possession of the property pairs ($\alpha,1$) and ($\beta,2$), and which is subsequently observed in possession of the property pairs ($\alpha,A$) and ($\beta,B$). Nothing, therefore, stands in the way of conceiving of the spatial relations which make up the formal aspects of quantum objects as reflexive relations entertained by a \emph{single} relatum, and to posit this as the \emph{sole} metaphysical substrate of \emph{all} properties measured or experienced. We shall call it ``the One.''

While the click of a detector is the most basic kind of outcome-indicating event, it is only a particular kind of genuinely fortuitous event. Every measurement is genuinely fortuitous not merely with regard to its specific outcome (which may in fact be predictable) but more fundamentally with regard to its \emph{occurrence}. There are no causally sufficient conditions for the success of an attempted measurement. The redundancy typically built into a measurement apparatus can maximize the likelihood of success, but it can never guarantee success.%
\footnote{This likelihood is not one of the probabilities that quantum mechanics serves to assign. Quantum mechanics probability assignments are based on the assumption of a successful outcome, which is why the probabilities associated with the possible outcomes of a measurement always add up to~1.}

What shall we make of the fortuitous nature of  outcome-indicating events? If in spite of their fortuitous character---i.e., undeterred by their lack of a cause \emph{in the experienced world}---we adhere to the principle of causality, we will have to acknowledge the existence \emph{beyond} this world of a causal agent responsible for each and every click in this world. We cannot, as Falkenburg suggests,%
\footnote{As part of her ``critical view of quantum reality,'' \citet[p.~259]{Falkenburg} has proposed ``a weakened version of the traditional metaphysics of substance,'' where ``the metaphysical carriers of the properties are cancelled'' but ``some metaphysical glue is left which makes [the properties] stick together.'' In spite of the fact that, ``[o]perationally, the particle behind a track is \emph{nothing but} the repeated localization of conserved dynamic quantities,'' she concludes that ``we cannot but interpret the repeatability as indicating an underlying entity'' (p.~260), which obviously will be a different one for different tracks.}
posit a unique causal agent behind the clicks that make up any particular track, except FAPP. What is warranted without qualification is to associate a single causal agent with every click (not only those that make up a particular track). And if we do, we will be able to identify the One---the sole substrate of all properties---also as the sole causal agent responsible for all clicks, and more generally for all successful measurements.

\section{Schr\"odinger's take}\label{sec:Schr-take}
So far the question of whether we only experience our respective private worlds or actually experience one and the same world was left open. But if the spatial relations that make up the formal aspect of quantum objects are reflexive and thus entertained by a \emph{single} relatum, or if there is a single substance underlying \emph{every} property measured or experienced, or if there is a single causal agent responsible for \emph{every} successful measurement, then there has to be more to the agreement between ``the content of any one sphere of consciousness and any other'' \citep[p.~69]{SchrWhatIsReal} than is warranted by language.

To \citet[p.~106]{SchrWhatIsReal}, the agreement of the content of my sphere of consciousness with the content of yours was not rationally comprehensible: ``In order to grasp it we are reduced to two irrational, mystical hypotheses.'' Previously we noted what he thought of ``the so-called hypothesis of the real external world.'' The second hypothesis, which he endorsed, was that ``we are all really only various aspects of the One.'' The multiplicity of minds, he wrote in another work \citep{SchrLifeMindMatterAP}, ``is only apparent, in truth there is only one mind. This is the doctrine of the Upanishads. And not only of the Upanishads.'' The Upanishads are ancient Sanskrit texts, which contain many central concepts and ideas of classical Indian philosophy.

The ``One'' mentioned by Schr\"odinger is the one Self of the Upanishads, the Ultimate Subject from which we are separated by a veil of self-oblivion. The same veil, according to the Upanishads, also prevents us from perceiving the Ultimate Object, as well as its fundamental identity with the Ultimate Subject. It prevents us from perceiving that the world is something that the One (qua Ultimate Object) manifests to itself (qua Ultimate Subject)---and therefore to us who are but ``various aspects of the One''.%
\footnote{If ``to Western thought this doctrine has little appeal,'' \citet[pp. 129--130]{SchrLifeMindMatterPO} remarks, it is because our science ``is based on objectivation, whereby it has cut itself off from an adequate understanding of the Subject of Cognizance, of the mind.'' To which he adds that ``this is precisely the point where our present way of thinking does need to be amended, perhaps by a bit of blood-transfusion from Eastern thought. That will not be easy, we must beware of blunders---blood-transfusion always needs great precaution to prevent clotting. We do not wish to lose the logical precision that our scientific thought has reached, and that is unparalleled anywhere at any epoch.''}

If at bottom we are all the same subject---without being aware of it, except by a genuinely mystical experience that is hard to come by---then we have to conceive of two poises of consciousness or modes of awareness, one in which the One manifests the world to itself \emph{aperspectivally}, as if experienced from no particular location or from everywhere at once, and one in which the One manifests the world to itself \emph{perspectivally}, as if experienced by a multitude of subjects from a multitude of locations.%
\footnote{An aperspectival consciousness features prominently in the respective works of Jean Gebser (\citeyear{GebserEPO}) and \citet{SA-TLD}. Such a consciousness transcends the distantiating viewpoint of our perspectival outlook. There, the subject is where its objects are; it knows them by identity, by \emph{being} them. The familiar dimensions of phenomenal space (viewer-centered depth and lateral extent) come into being in a secondary poise, in which the One views the world in perspective. There, objects are seen from ``outside,'' as presenting their surfaces. Concurrently, the dichotomy between subject and object becomes a reality, for a subject identified with an individual form cannot be overtly identical with the substance that constitutes all forms.}
In the former poise, the One contains the world; in the latter, it is contained in it. The fact that reality is relative to consciousness---whether unitary and aperspectival or multiply situated and perspectival---takes the edge of Husserl's paradox of human subjectivity. As situated aspects of the One, we appear to be objects in the world; as the One, we both constitute and contain the world. 

In the view of the Upanishads, all knowledge, all experience is founded on identity. What ultimately exists, independently of anything else, is indistinguishably (i)~a consciousness that contains, (ii)~a substance that constitutes, and (iii)~an infinite quality and delight (\emph{\={a}nanda}) that experiences and expresses itself in form and movement. If the One is essentially this infinite, self-existent quality and delight, one understands why the One would adopt a multitude of standpoints within the world that it manifests to itself: a mutual creative self-experience offers a greater variety of delight than a solitary one. If the One adopts a multitude of localized standpoints, knowledge by identity takes the form of direct knowledge: each individual knows the others directly, without mediating representations. If the One identifies itself with each particular form or standpoint \emph{to the exclusion of} all others, knowledge of other forms will be reduced to an indirect knowledge, i.e., a direct knowledge by the individual of some of its own attributes (think electrochemical pulses in a brain), which serve as representations of the other forms. 

\section{The cognitive synergy of the One}\label{sec:synergy}
Perhaps the most pressing issue in the philosophy of mind concerns what is known as ``intentionality'': how can anything in the world, such as a neural firing pattern, represent---and, more generally, refer to or be about---anything else in the world?%
\footnote{In recent years, many philosophers have put a high priority on providing a reductionist account of intentional categories, such as beliefs and desires. If this project were to fail, \citet{Fodor87} opines, ``that would be, beyond comparison, the greatest intellectual catastrophe in the history of our species; if we're that wrong about the mind, then that's the wrongest we've ever been about anything. The collapse of the supernatural, for example, didn't compare; theism never came close to being as intimately involved in our thought and our practice---especially our practice---as belief/desire explanation is. Nothing except, perhaps, our commonsense physics---our intuitive commitment to a world of observer-independent, middle-sized objects---comes as near our cognitive core as intentional explanation does.''}
An idea that has been dominant throughout history is that representations are encodings. Much has been learned about the processes by which the brain extracts information from the images falling on the retinas \citep{Hubel95,Enns04}. This information is said to be encoded in patterns of electrochemical pulses, and if these patterns are to be experienced as (or to give rise to experiences of) a world extended in space and time, they have to be \textit{decoded} or \emph{interpreted}. The decoding or interpretation presupposes, \emph{inter alia}, acquaintance with the expanse of space and the passing of time, and such acquaintance is not something that neural processes can provide.%
\footnote{Like the color of a Burmese ruby, spatial extension is a quality that can only be defined by ostentation---by drawing attention to something of which we are directly aware. If you are not convinced, try to explain to my friend Andy, who lives in a spaceless world, what space is like. Andy is good at math, so he understands you perfectly if you tell him that space is like a set of all triplets of real numbers. But if you believe that this gives him a sense of the expanse we call space, you are deluding yourself. \emph{We} can imagine triplets of real numbers as points embedded in space; he cannot. \emph{We} can interpret the difference between two numbers as the distance between two points; he cannot. At any rate, he cannot associate with the word ``distance'' the phenomenal remoteness it conveys to us. And much the same goes for time. Time passes, and the only way to know this is to be aware of it.}

\begin{figure}[t]\begin{center}
\includegraphics[width=3in]{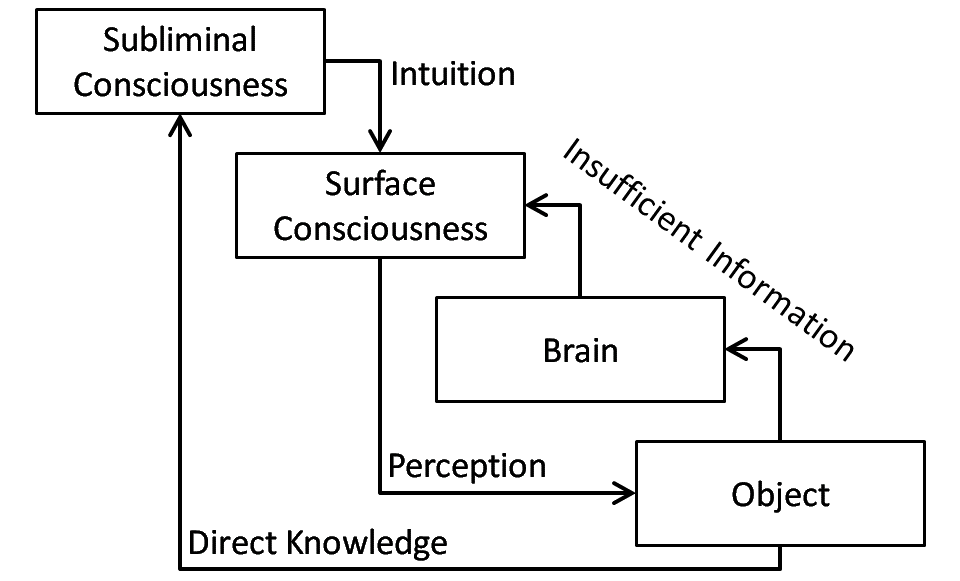}
\caption{Our indirect (hence representative) knowledge is the meeting point of information flowing inward from the external object and information flowing outward from a subliminal source.}\label{fig.subliminal}
\end{center}\end{figure}

Some, like \citet{BickhardRichie}, noting that encodings always require an interpreter, are intent on repudiating the interpreter. What is really required, however, is an understanding of the nature of the interpreter and the basis of its interpretive capacity, and this the Upanishadic theory of existence can provide. According to it, the incomplete quantitative information provided by neural firing patterns is supplemented by a qualitative direct knowledge that springs from a subliminal source, as Fig.~\ref{fig.subliminal} suggests. Our indirect knowledge would not be possible if it were not supported by a subliminal direct knowledge, even as direct knowledge would not be possible if it were not founded on identity. In the words of \citet[pp. 560--61]{SA-TLD},
\bq
In the surface consciousness knowledge represents itself as a truth seen from outside, thrown on us from the object, or as a response to its touch on the sense, a perceptive reproduction of its objective actuality\dots. Since it is unable to \dots\ observe the process of the knowledge coming from within, it has no choice but to accept what it does see, the external object, as the cause of its knowledge\dots. In fact, it is a hidden deeper response to the contact, a response coming from within that throws up from there an inner knowledge of the object, the object being itself part of our larger self.
\eq

\section{Evolution: the Upanishadic view}\label{sec:evolution}
I suppose we can all appreciate the advantage of an ontology that has at its core an infinite Quality/Delight over a framework of thought according to which what is ultimately real is a multitude of entities (fundamental particles or spacetime points) lacking intrinsic quality or value. In many traditions such a multiplicity is fittingly referred to as ``dust.'' But a questions remains: why should a self-existent and infinite conscious Quality and Delight not only adopt a multitude of standpoints but also identify itself with each to the apparent exclusion of the others? The answer lies in the fact that the main plot of the self-manifestation of the One is evolution, or rather a cycle of self-concealment and self-discovery. 

From the point of view of the Upanishads, evolution presupposes \emph{involution}. Involution consists in a stepwise departure from the original status of the One as at once an all-containing consciousness, an all-constituting substance, and an infinite quality/value/delight. The first step towards involution is individuation. One nice thing about the process of individuation is that we can feel as if we understand it. We all know first-hand what it means to imagine things. So we can easily conceive of a consciousness that creates its own content. With a little effort we can also conceive of consciousness as simultaneously adopting a multitude of standpoints, and of some or all of its creative activity as simultaneously proceeding from these several standpoints. We also know first-hand the phenomenon of exclusive concentration, when awareness is focused on a single object or task, while other goings-on are registered, and other tasks attended to, subconsciously (if at all). 

Further possibilities present themselves. The multiple concentration of consciousness may be inclusive, or it may be exclusive. Involution begins in earnest when the individual subjects lose sight of their mutual identity and, as a result, lose access to the aperspectival view of things. Carried further, involution renders consciousness implicit in its aspect of formative force. Carried still further, it renders formative force implicit in inanimate forms. And carried to its furthest extreme, it renders the principle of form implicit in a multitude of formless entities. And since these formless entities are indistinguishable and therefore (by the principle of the identity of indiscernibles) numerically identical, involution ends with the One effectively deprived of its innate consciousness and self-determining force---the Ultimate Subject, infinitely creative, rendered implicit in the Ultimate Object. This (or something much like it) is how the stage for the adventure of evolution was set.

We have now strayed, inevitably, into the vexing territory of theodicy: what could justify this adventure, considering all the pain and suffering that (in hindsight) it entails? Certainly not an extra-cosmic Creator imposing these evils on his creatures. But the One of the Upanishads is no such monster; it imposes these things \emph{on itself}. But still---why? Imagine for a moment that you are all-powerful and all-knowing. Could you experience the joy of winning a victory, of overcoming difficulties and oppositions, of making discoveries, of being surprised? You could not.
To make all of this possible, you impose limitations on your inherent power and knowledge. In the words of Sri Aurobindo:
\bq
a play of self-concealing and self-finding is one of the most strenuous joys that conscious being can give to itself, a play of extreme attractiveness. There is no greater pleasure for man himself than a victory which is in its very principle a conquest over difficulties, a victory in knowledge, a victory in power, a victory in creation over the impossibilities of creation\dots. There is an attraction in ignorance itself because it provides us with the joy of discovery, the surprise of new and unforeseen creation\dots. If delight of existence be the secret of creation, this too is one delight of existence; it can be regarded as the reason or at least one reason of this apparently paradoxical and contrary Lila. \cite[pp. 426--27]{SA-TLD}
\eq
\emph{L\={\i}l\={a}} is a term of Indian philosophy which describes the manifested world as the field for a joyful sporting game made possible by self-imposed limitations.

\section{Concluding remarks: physics, free will, and the future of evolution}\label{sec:future}
So how does physics fit into this scheme of things? If the force at work in the world is an \emph{infinite} force working under self-imposed constraints, we can stop being spooked by the quantum-mechanical correlation laws.%
\footnote{There is an ever-growing number of ``no-go'' theorems \citep{Bell64,Bell66,KS,GHZ,Klya} demonstrating the impossibility of dynamical explanations of the correlations that quantum mechanics successfully predicts.} 
There is no need for mechanistic or naturalistic explanations of the working of an infinite force. What we need to ask is why the force at work in the world works under constraints, which question we have answered, and why these constraints have the particular form that they do. Briefly, a world about which anything coherent can be said requires sufficiently stable, re-identifiable forms. If these are to be manifested by means of spatial relations between relata that lack spatial extent, the spatial relations, as well as the corresponding relative momenta, must be indefinite, uncertainty relations must hold, the relata must be fermions---in short, something very much like quantum mechanics must hold~\citep{JustSo,Mohrhoff-QMexplained}. And if the manifested world is, in addition, to contain individuals capable of telling stories about sufficiently stable, re-identifiable objects, the standard model and general relativity must hold at least as effective theories \citep[Chap.~25]{Mohrhoff-book}.

A force working under self-imposed constraints is also capable of lifting its constraints. Their purpose was to set the stage for the drama of evolution, not to direct the drama. We are actors in this drama, and if our free will is limited, it is not because it is illusory. Needless to say, there is but one way in which complete freedom can be attained, and that is by becoming the sole determinant of the goings-on in the world. We are in possession of true freedom to the extent that we are not only consciously but also dynamically identified with the One. Absent this identification, our sense of being the proud owner of a libertarian free will is a misappropriation of a power which belongs to our subliminal self, and which often works towards goals that are at variance with our conscious intentions.

The first aim of the force at work in the world---after setting the stage for evolution---is to bring into play the principles of life and mind. Because it has to accomplish this through tightly constrained modifications of the initial laws \citep{Mohrhoff-JCS},%
\footnote{The reason we lack direct evidence of these modifications is the Houdiniesque nature of this manifestation. ``If delight of existence be the secret of creation,'' there have to be serious limitations on the extent of possible modifications. Given the means at our disposal, it will therefore be virtually impossible to discern where and when they occur.}
the evolution of life necessitates the creation of increasingly complex organisms, and the evolution of mind necessitates the creation of increasingly complex nervous systems. This explains why the representations that mediate indirect knowledge require something of the order of a hundred billion neurons.

Evolution is not finished. When life emerged, what essentially emerged was the power to execute creative ideas. When mind (or consciousness as we know it) emerged, what essentially emerged was the power to generate such ideas. The true nature of these principles is obscured by the fact that the requisite anatomy must first be established, and the more pressing tasks of self-preservation and self-replication must first be attended to. What has yet to emerge is the power to develop into expressive ideas the infinite Quality at the heart of reality. When this happens, the entire creative process---i.e., the development of Quality into Form using mind to generate expressive ideas and life to execute them---will be conscious and deliberate. 

If all of this sounds phantasmagoric, it is in large part because our theoretical dealings with the world are conditioned by the manner in which we, at this point in history, experience the world. We conceive of the evolution of consciousness, if not as a sudden lighting up of the bulb of sentience, then as a progressive emergence of ways of experiencing a world that exists independently of being experienced. 
There is no such world. There are only different ways in which the One manifests the world to itself. These different ways have been painstakingly documented by Jean \citet{GebserEPO}. One characteristic of the ``structures of consciousness'' that have successively emerged or are on the verge of emerging is their dimensionality. An increase in the dimensionality of the consciousness to which the world is manifested is tantamount to an increase in the dimensionality of the manifested world. 

Consider, by way of example, the consciousness structure that immediately preceded the present and still dominant one. One of its characteristics is the notion that the world is enclosed in a sphere, with the fixed stars attached to its boundary, the firmament. \emph{We} cannot but ask: what is beyond that sphere? Those who held this notion could not, because for them the third dimension of space---viewer-centered depth---did not at all have the reality it has for us. Lacking our sense of this dimension, the world experienced by them was in an important sense two-dimensional. This is why they could not handle perspective in drawing and painting, and why they were unable to arrive at the subject-free ``view from nowhere,'' which is a prerequisite of modern science. All this became possible with the consolidation, during the Renaissance, of our characteristically three-dimensional consciousness structure.

Our very concepts of space, time, and matter are bound up with our present consciousness structure. This made it possible to integrate the location-bound outlook of a characteristically two-dimensional consciousness into an effectively subject-free world of three-dimensional objects. Matter as we know it was the result. It is not matter that has created consciousness; it is consciousness that has created matter, first by its self-concealment, or involution, in an apparent multitude of formless particles, and again by evolving our present mode of experiencing the world. Ahead lies the evolution of a consciousness structure---and thereby of a world---that transcends our time- and space-bound perspectives. Just as the mythological thinking of the previous consciousness structure could not foresee the technological explosion made possible by science, so science-based thinking cannot foresee the consequences of the birth of a new world, brought about, not by technological means, but by a further increase in the dimensionality of consciousness.

\section*{Appendix: A minimal critique of QBism}
As \citet[p.~339]{Falkenburg} acknowledges, ``Bohr’s complementarity view of quantum mechanics already suggested long ago a way of avoiding the dilemma of either empiricism or metaphysical realism.'' If, as she notes, ``no unambiguous philosophical interpretation of quantum theory emerged out of them,'' it was because nobody seemed to realize or take seriously (i)~a crucial presupposition of Bohr's philosophical reflections---to wit, ``the subjective character of all experience'' \citep[p.~259]{BCW7}---and (ii)~a crucial implication of the same, which is that the classical environment owes its definiteness solely to the incontestable definiteness of conscious experience \citep{UJMBohrObIrrev}. 

While, to my mind, the greatest merit of QBism lies in its spirited defense of these crucial points, its most significant shortcoming lies in its failure to recognize that in constructing a common body of reality, the users (of quantum mechanics) or agents (in a quantum world) automatically \emph{objectivize} the intrinsic definiteness of their private experiences. When Fuchs \&\ friends state that the only thing a QBist ``does not model with quantum mechanics is her own direct internal awareness of her own private experience'' \citep{FMS2014}, so that QBism must treat ``all physical systems in the same way, including atoms, beam splitters, Stern-Gerlach magnets, preparation devices, measurement apparatuses, all the way to living beings and other agents'' \citep{FS2015}, they ignore this obvious fact. If quantum mechanics is a general theory of action and response, according to which every action (on the external world) counts as a measurement and every response (by the external world) counts as the outcome of a measurement, as \citet{FS2004} claim, it must be possible to put into words all possible actions along with their possible responses. And for this it must be possible to talk about beam splitters, Stern-Gerlach magnets, preparation devices, etc., in classical terms.

As \citet{Fuchs_Notwithstanding} sees it, a measurement (i.e., an action taken to elicit one of a set of possible experiences) ``can be anything from running across the street at L'\'{E}toile in Paris (and gambling upon one's life) to a sophisticated quantum information experiment (and gambling on the violation of a Bell inequality).'' Missing the distinction between a domain that is directly accessible to sensory experience, and a domain that is speakable only contextually, in terms of quantum phenomena, QBism also fails to make the necessary distinction between two kinds of action: (i)~actions by which we reach into the quantum domain, whose predictable outcomes are linked to their known initial and boundary conditions in accordance with the Born rule, and (ii)~actions whose predictable outcomes are linked to their known initial and boundary conditions in accordance with the classical law of total probability. Running across the street at L'\'{E}toile in Paris (and gambling upon one's life) obviously belongs to the latter kind.

Everything we believe---including everything we claim to know---is a belief. QBists  are absolutely right about this, and I second their interpretation of probability as a personal degree of belief.%
\footnote{Specifically, I am with them in rejecting the so-called eigenvalue--eigenstate link, according to which the assignment of probability~1 to the possible value~$q$ of an observable~$Q$ justifies the belief that $Q$ actually \emph{has} the value~$q$.}
The objective world---i.e., ``the common external world we have all negotiated with each other'' \citep{MerminQBnotCop}---is what we collectively believe to exist. The implicit assumption underlying this collective belief is not only that the spatiotemporal structure of my perceptual awareness and the logical structure of my thought are the same as the spatiotemporal structure of your perceptual awareness and the logical structure of your thought: we share the cognitive structures to which our basic physical concepts owe their meanings. The implicit assumption is also that my experiences are as definite as yours. Wigner may be ignorant of the outcome experienced by his friend, but he cannot be cavalier about the definiteness of her experiences. It simply makes no sense to describe the state of Wigner's friend \citep{Wigner61} as a coherent superposition of having experienced two different outcomes---a possibility envisioned by \citet{Fuchs_Notwithstanding}. By the very nature of our common external world, or by the manner in which it is constructed by us, Wigner is not merely justified but \emph{required} to assign to the state of his friend an incoherent mixture reflecting his ignorance of her actual experience. And the same applies to the respective states assigned by Alice to Bob and by Bob to Alice in experiments with spin-$1/2$ particles prepared in the singlet state, contrary to what is claimed by \citet{FMS2014}.

\bigskip\noindent\textbf{Ulrich J. Mohrhoff} studied physics at the University of G\"ottingen and at the Indian Institute of Science in Bangalore. He teaches at the Sri Aurobindo International Centre of Education (Pondicherry, India) and is the author of the textbook \emph{The World According to Quantum Mechanics: Why the Laws of Physics Make Perfect Sense After All} (Singapore: World Scientific, Second Edition 2018). His research interests are the Foundations of Quantum Physics, the Philosophy of Mind, and the Interface between Contemporary Physics and Indian Philosophy.

\medskip\noindent The author affirms that there is no conflict of interest.

\begin{thebibliography}{00}
\bibitem[Bell(1964)]{Bell64}
Bell, J.S. (1964). On the Einstein Podolsky Rosen paradox. \textit{Physics Physique Fizika}, \textit{1}, 195--200.

\bibitem[Bell(1966)]{Bell66}
Bell, J.S. (1966). On the problem of hidden variables in quantum mechanics. \textit{\mbox{Reviews} of Modern Physics}, \textit{38}, 447--452.

\bibitem[Bell(1990)]{Bell90}
Bell, J.S. (1990). Against ``measurement.'' In A.I. Miller (Ed.), \textit{62 Years of Uncertainty}, pp. 17--31. New York: Plenum.

\bibitem[Bickhard and Richie(1983)]{BickhardRichie}
Bickhard, M.H. \& Richie, D.M. (1983). \textit{On the Nature of Representation}. New York: Praeger.

\bibitem[Bitbol(2007)]{BitbolQMCR}
Bitbol, M. (2007). Schr\"odinger against particles and quantum jumps. In J. Evans \& A.S. Thorndike (Eds.), \textit{Quantum Mechanics at the Crossroads}, pp. 81--106. Berlin: Springer.

\bibitem[Bitbol(2010)]{Bitbol2010}
Bitbol, M. (2010). Reflective metaphysics: Understanding quantum mechanics from a Kantian standpoint. \textit{Philosophica}, \textit{83}, 53--83.

\bibitem[Bitbol et al.(2009)]{Bitboletal}
Bitbol, M., Kerszberg, P. \& Petitot, J. (Eds.) (2009). \textit{Constituting Objectivity: Transcendental Perspectives on Modern Physics}. Springer Science+Business Media.

\bibitem[Bohr(1985)]{BCW6}
Bohr, N. (1985). \textit{Niels Bohr: Collected Works}, Vol. 6. Amsterdam: North-Holland.

\bibitem[Bohr(1996)]{BCW7}
Bohr, N. (1996). \textit{Niels Bohr: Collected Works}, Vol. 7. Amsterdam: Elsevier.

\bibitem[Bohr(1999)]{BCW10}
Bohr, N. (1999). \textit{Niels Bohr: Collected Works}, Vol. 10. Amsterdam: Elsevier.

\bibitem[Boscovich(1922)]{Bosco}
Boscovich, R. (1922). \textit{A theory of Natural Philosophy}. Cambridge, MA:  MIT Press.

\bibitem[Brock(2009)]{Brock}
Brock, S. (2009). Old wine enriched in new bottles: Kantian flavors in Bohr's viewpoint of complementarity. In \cite{Bitboletal}, pp. 301--316.

\bibitem[Busch et al.(1996)]{BLM96} 
Busch, P., Lahti, P.J. \&  Mittelstaedt, P. (1996). \textit{The Quantum Theory of Measurement}. Berlin: Springer.

\bibitem[Caves et al.(2002)]{CFS2002}
Caves, C.M., Fuchs, C.A., Schack, R.  (2002). Quantum probabilities as Bayesian probabilities. \textit{Physical Review A}, \textit{65}, 022305.

\bibitem[Chevalley(1994)]{Chevalley94}
Chevalley, C. (1994). Niels Bohr's words and the Atlantis of Kantianism. In \cite{FF94}, pp. 33--55.

\bibitem[Cuffaro(2010)]{Cuffaro}
Cuffaro, M.E. (2010). The Kantian framework of complementarity. \textit{Studies in History and Philosophy of Modern Physics}, \textit{41}, 309--317.

\bibitem[Dennett and Steglich-Petersen(2008)]{hilary}
Dennett, D. \& Steglich-Petersen, A. (Eds.) (2008). \textit{The Philosophical Lexicon}, http://www.philosophicallexicon.com/.

\bibitem[Enns(2004)]{Enns04}
Enns, J.T. (2004). \textit{The Thinking Eye, the Seeing Brain: Explorations in Visual Cognition}. New York: Norton \& Company.

\bibitem[Falkenburg(2007)]{Falkenburg}
Falkenburg, B. (2007). \textit{Particle Metaphysics: A Critical Account of Subatomic Reality}. Berlin: Springer.

\bibitem[Falkenburg(2009)]{Falkenburg2009}
Falkenburg, B. (2009). A critical account of physical reality. In \cite{Bitboletal}, pp. 229--248.

\bibitem[Faye and Folse(1994)]{FF94}
Faye, J., Folse \& H.J. (Eds.) (1994). \textit{Niels Bohr and Contemporary Philosophy}. Dordrecht: Kluwer.

\bibitem[Fodor(1987)]{Fodor87}
Fodor, J. (1987). \textit{Psychosemantics: The problem of meaning in the philosophy of mind}, p.~xii. Cambridge, MA: Bradford Books/MIT Press.

\bibitem[Folse(1994)]{Folse94}
Folse, H. (1994). Bohr's framework of complementarity and the realism debate. In \cite{FF94}, pp. 119--139.

\bibitem[Van Fraassen(1990)]{vF1990}
Fraassen, B.C. van (1990). The problem of measurement in quantum mechanics. In P. Lahti \& P Mittelstaedt (Eds.), \textit{Symposium on the Foundations of Modern Physics}, pp. 497--503. Singapore: World Scientific.

\bibitem[Fuchs(2010)]{Fuchs2010} 
Fuchs, C.A. (2010). Quantum Bayesianism at the Perimeter. \textit{Physics in Canada}, \textit{66} (2), 77--81.

\bibitem[Fuchs(2017)]{Fuchs_Notwithstanding}
Fuchs, C.A. (2017). Notwithstanding Bohr, the reasons for QBism. \textit{Mind and Matter}, \textit{15} (2), 245--300.

\bibitem[Fuchs and Schack(2004)]{FS2004}
Fuchs, C.A. \& Schack, R. (2004). Unknown quantum states and operations: a Bayesian view. In M. Paris \& J. \v{R}eh\'{a}\v{c}ek (Eds.), \textit{Quantum State Estimation}, pp. 147--187. Berlin: Springer.

\bibitem[Fuchs and Schack(2015)]{FS2015}
Fuchs, C.A. \& Schack, R. (2015). QBism and the Greeks: Why a quantum state does not represent an element of physical reality. \textit{Physica Scripta}, \textit{90} (1), 015104.

\bibitem[Fuchs et al.(2014)]{FMS2014}
Fuchs, C.A., Mermin, N.D. \& Schack, R. (2014). An introduction to QBism with an application to the locality of quantum mechanics.  \textit{American Journal of Physics}, \textit{82},  749--754.

\bibitem[Gebser(1986)]{GebserEPO}
Gebser, J. (1986). \textit{The Ever-Present Origin}. Athens, OH: Ohio University Press.

\bibitem[Greenberger et al.(1989)]{GHZ}
Greenberger, D.M., Horne, M. \& Zeilinger, A. (1989). Going beyond Bell's theorem. In M. Kafatos (Ed.), \textit{Bell's Theorem, Quantum Theory, and Conceptions of the Universe}, pp. 69--72. Dordrecht: Kluwer.

\bibitem[Grupen and Shwartz(2008)]{GS}
Grupen, C. \& Shwartz, B. (2008). \textit{Particle Detectors}, 2nd Edition, Chap.~9. Cambridge, UK: Cambridge University Press.

\bibitem[Heisenberg(1927)]{HeisenbergBahn}
Heisenberg, W. (1927). \"{U}ber den anschaulichen Inhalt der quan\-ten\-the\-o\-re\-ti\-schen Kinematik und Mechanik. \textit{Zeitschrift f\"{u}r Physik}, \textit{43}, 172--198.

\bibitem[Honner(1982)]{Honner1982}
Honner, J. (1982). The transcendental philosophy of Niels Bohr. \textit{Studies in History and Philosophy of Science A}, \textit{13}, 1--29.

\bibitem[Hooker(1994)]{Hooker94}
Hooker, C.A. (1994). Bohr and the crisis of empirical intelligibility: An essay on the depth of Bohr's thought and our philosophical ignorance. In \cite{FF94}, pp. 155--199.

\bibitem[Hubel(1995)]{Hubel95}
Hubel, D.H. (1995). \textit{Eye, Brain, and Vision}.  New York: Scientific American Library.

\bibitem[Husserl(1970)]{Husserl}
Husserl, E. (1970). \textit{The Crisis of European Sciences and Transcendental Phenomenology}.  Evanston, IL: Northwestern University Press.

\bibitem[James(2004)]{James04}
James, I. (2004). \textit{Remarkable Physicists From Galileo to Yukawa}. Cambridge, UK: Cambridge University Press.

\bibitem[Joos et al.(2003)]{Joosetal2003}
Joos, E., Zeh, H.D., Kiefer, C., Giulini, D.J.W., Kupsch, J. \& Stamatescu, \mbox{I.-O.} (2003). \textit{Decoherence and the Appearance of a Classical World in Quantum Theory}, 2nd Edition. New York: Springer.

\bibitem[Kaiser(1992)]{Kaiser1992}
Kaiser, D. (1992). More roots of complementarity: Kantian aspects and influences.  \textit{Studies in History and Philosophy of Science A}, \textit{23}, 213--239.

\bibitem[Kant(1998)]{KantCPR}
Kant, I. (1998). \emph{Critique of Pure Reason}. Cambridge, UK: Cambridge University Press.

\bibitem[Ketterle(2003)]{Ketterle}
Ketterle, W. (2003). Talk given at the annual meeting of the German Physical Society in Hannover.

\bibitem[Klyachko et al.(2008)]{Klya}
Klyachko, A.A., Can, M.A., Binicio\v{g}lu, S. \& Shumovsky, A.S. (2008). A simple test for hidden variables in the spin-1 system. \textit{Physical Review Letters}, \textit{101}, 020403.

\bibitem[Kochen and Specker(1967)]{KS}
Kochen, S. \& Specker, E. (1967). The problem of hidden variables in quantum mechanics. \textit{Journal of Mathematics and Mechanics}, \textit{17}, 59--87.

\bibitem[Ladyman and Ross(2007)]{LadyRoss}
Ladyman, J. \& Ross, D. (with Spurrett, D. \& Collier, J. (2007). \textit{Every Thing Must Go: Metaphysics Naturalized}. Oxford: Oxford University Press.

\bibitem[Levine(2001)]{Levine}
Levine, J. (2001). \textit{Purple Haze: The Puzzle of Consciousness}, p.~78. Oxford: Oxford University Press.

\bibitem[Locke(1997)]{Locke}
Locke, J. (1997). \textit{An Essay Concerning Human Understanding}, Book II, Chapter XXIII, 2--3. London: Penguin Books.

\bibitem[London and Bauer(1939/1983)]{LonBau}
London, F.W. \& Bauer, E. (1939/1983). \textit{La Th\'eorie de I'observation en m\'ecanique quantique}. Paris: Hermann; Trans. in J.A. Wheeler \& W.H. Zurek (Eds.)  \textit{Quantum Theory and Measurement}, pp. 217--259. Princeton, NJ: Princeton University Press.

\bibitem[MacKinnon(2012)]{MacKinnon2012}
MacKinnon, E. (2012). \textit{Interpreting Physics: Language and the Classical/Quantum Divide}. Springer Science+Business Media.

\bibitem[Mermin(1998)]{Mermin98}
Mermin, N.D. (1998). What is quantum mechanics trying to tell us? \textit{American Journal of Physics}, \textit{66}, 753--767.

\bibitem[Mermin(2014)]{MerminNatureQBism}
Mermin, N.D. (2014). QBism puts the scientist back into science. \textit{Nature}, \textit{507}, 421--423.

\bibitem[Mermin(2017)]{MerminQBnotCop}
Mermin, N.D. (2017). Why QBism is not the Copenhagen interpretation and what John Bell might have thought of it. In R. Bertlmann \& A. Zeilinger (Eds.), \textit{Quantum [Un]Speakables II: 50 Years of Bell's Theorem}, pp. 83--93. Berlin: Springer.

\bibitem[Mermin(2019)]{Mermin2019}
Mermin, N.D. (2019). Making better sense of quantum mechanics. \textit{Reports on Progress in Physics}, \textit{82}, 012002.

\bibitem[Misner et al.(1973)]{Misneretal}
Misner, C.W., Thorne, K.S. \& Wheeler, J.A. (1973). \textit{Gravitation}, p. 1215. San Francisco, CA: W.H. Freeman and Company.

\bibitem[Mittelstaedt(1998)]{Mittelstaedt98} 
Mittelstaedt, P. (1998). \textit{The Interpretation of Quantum Mechanics and the Measurement Process}. Cambridge, UK: Cambridge University Press.

\bibitem[Mohrhoff(1999)]{Mohrhoff-JCS}
Mohrhoff, U. (1999). The physics of interactionism. In B. Libet, A. Freeman \& K. Sutherland (Eds.), \textit{The Volitional Brain}, pp. 165--184. Exeter, UK: Imprint Academic.

\bibitem[Mohrhoff(2000)]{Mohrhoff2000}
Mohrhoff, U. (2000). What quantum mechanics is trying to tell us. \textit{American Journal of Physics}, \textit{68}, 728--745.

\bibitem[Mohrhoff(2002)]{JustSo} 
Mohrhoff,  U. (2002). Why the laws of physics are just so. \textit{Foundations of Physics} \textit{32} (8), 1313--1324.

\bibitem[Mohrhoff(2009)]{Mohrhoff-QMexplained}
Mohrhoff, U. (2009). Quantum mechanics explained. \textit{International Journal of Quantum Information}, \textit{7}, 435--458.

\bibitem[Mohrhoff(2014)]{UJM-QBismAppraisal}
Mohrhoff, U. (2014). QBism: A critical appraisal. arXiv:1409.3312 [quant-ph].

\bibitem[Mohrhoff(2017)]{Mohrhoff2017}
Mohrhoff, U.  (2017). Quantum mechanics in a new light. \textit{Foundations of Science}, \textit{22}, 517--537.

\bibitem[Mohrhoff(2018)]{Mohrhoff-book}
Mohrhoff, U. (2018). \textit{The World According to Quantum Mechanics: Why the Laws of Physics Make Perfect Sense After All}, 2nd Edition. Singapore: World Scientific.

\bibitem[Mohrhoff(2019)]{UJMBohrObIrrev}
Mohrhoff, U. (2019). Niels Bohr, objectivity, and the irreversibility of measurements. \textit{Quantum Studies: Mathematics and Foundations}. https://doi.org/10.1007/s40509-019-00213-6.

\bibitem[Mohrhoff(2021)]{Mohrhoff-B4B}
Mohrhoff, U. (2021). ``B'' is for Bohr. To appear in D. Aerts, J. Arenhart, C. de Ronde \& G. Sergioli (Eds.), \textit{Probing the Meaning and Structure of Quantum Mechanics} (Vol. 3). Singapore: World Scientific; arXiv:1905.07118 [quant-ph].

\bibitem[Neumann(1932/1955)]{vN}
Neumann, J. von (1932/1955). \textit{Mathematische Grundlagen der Quantenmechanik}. Berlin: Springer; \textit{Mathematical Foundations of Quantum Mechanics}. Princeton, NJ: Princeton University Press.

\bibitem[Newton(1718)]{Newton-nature}
Newton, I. (1718). \textit{Opticks: A Treatise of the Reflections, Refractions, Inflexions and Colours of Light}, Second Edition. http://www.newtonproject.ox.ac.uk/\break view/texts/normalized/NATP00051.

\bibitem[Nietzsche(2002)]{NietzscheBGE}
Nietzsche, F.W. (2002). \textit{Beyond Good and Evil}. Cambridge, UK: Cambridge University Press.

\bibitem[Pitowsky(2006)]{Pitowsky2006} 
Pitowsky, I. (2006). Quantum mechanics as a theory of probability. In W.~Demopoulos and I.~Pitowsky (Eds.), \textit{Physical Theory and Its Interpretation: Essays in Honor of Jeffrey Bub}, pp. 213--239. Dordrecht: Springer.

\bibitem[Poincar\'e(1905)]{Poincare}
Poincar\'e, H. (1905). \textit{Science and Hypothesis}, p.~50. London: The Walter Scott Publishing Company. 

\bibitem[Putnam(1987)]{Putnam_MFR}
Putnam, H. (1987). \textit{The Many Faces of Realism}. LaSalle, IL: Open Court.

\bibitem[Russell(1945)]{RussellHistory}
Russell, B. (1945). \textit{A History of Western Philosophy}, p.~202. New York: Simon and Schuster.

\bibitem[Schilling(1958)]{Schilling}
Schilling, H.K. (1958). A Human Enterprise: Science as lived by its practitioners bears but little resemblance to science as described in print.  \textit{Science}, \textit{127}, 1324--1327.

\bibitem[Sch\"onfeld and Thompson(2019)]{KantSEP}
Sch\"onfeld, M. \& Thompson, M. (2019). Kant's Philosophical Development. In E.N. Zalta (Ed.), \textit{The Stanford Encyclopedia of Philosophy} (Winter 2019 Edition). https://plato.stanford.edu/archives/win2019/entries/kant-development.

\bibitem[Schr\"odinger(1964)]{SchrWhatIsReal}
Schr\"odinger, E. (1964). What is real? In \textit{My View of the World}. Cambridge, UK: Cambridge University Press.

\bibitem[Schr\"odinger(1992a)]{SchrLifeMindMatterAP}
Schr\"odinger, E. (1992a). The Arithmetical Paradox. In \textit{What Is Life? With: Mind and Matter \&\ Autobiographical Sketches}, pp. 128--139. Cambridge, UK: Cambridge University Press.

\bibitem[Schr\"odinger(1992b)]{SchrLifeMindMatterPO}
Schr\"odinger, E. (1992b). The Principle of Objectivation. In \textit{What Is Life? With: Mind and Matter \&\ Autobiographical Sketches}, pp. 117--127. Cambridge, UK: Cambridge University Press.

\bibitem[Schr\"odinger(2014)]{SchrNGSH}
Schr\"odinger, E. (2014). \textit{Nature and the Greeks and Science and Humanism}. Cambridge, UK: Canto Classics.

\bibitem[Searle(2004)]{Searle2004}
Searle, J.R. (2004). \textit{Mind: A Brief Introduction}. New York: Oxford University Press.

\bibitem[Sri Aurobindo(2001)]{SA-Kena}
Sri Aurobindo (2001). \textit{Kena and other Upanishads}. Pondicherry, India: Sri Aurobindo Publication Department. https://bit.ly/SriAurobindo-Kena.

\bibitem[Sri Aurobindo(2003)]{SA-Isha}
Sri Aurobindo (2003). \textit{Isha Upanishad}. Pondicherry, India: Sri Aurobindo Publication Department. https://bit.ly/SriAurobindo-Isha.

\bibitem[Sri Aurobindo(2005)]{SA-TLD}
Sri Aurobindo (2005). \textit{The Life Divine}. Pondicherry, India: Sri Aurobindo Publication Department. https://bit.ly/SriAurobindo-TLD.

\bibitem[Ulfbeck and Bohr(2001)]{UlfBohr}
Ulfbeck, O. \& Bohr, A. (2001). Genuine Fortuitousness. Where Did That Click Come From? \textit{Foundations of Physics}, \textit{31}, 757--774.

\bibitem[Wallace(2008)]{Wallace2008}
Wallace, D. (2008). Philosophy of Quantum Mechanics. In D. Rickles (Ed.), \textit{The Ashgate Companion to Contemporary Philosophy of Physics}, pp. 16--98. Burlington, VT: Ashgate.

\bibitem[Von Weizs\"acker(1980)]{vW}
Weizs\"acker, C.F. von (1980). \textit{The Unity of Nature}, p.~328. New York: Farrar, Straus, and Giroux.

\bibitem[Wigner(1961)]{Wigner61}
Wigner, E.P. (1961). Remarks on the mind--body question. In I.J. Good (Ed.), \textit{The Scientist Speculates}, pp. 284--302. London: Heinemann.

\end{thebibliography}
\end{document}